\documentclass[preprint2]{aastex61}
\usepackage{amsmath}
\usepackage{amssymb}

\begin{document}

\title{Convective overshooting in \MakeLowercase{sd}B stars using the $\MakeLowercase k$-$\omega$ model}

\correspondingauthor{Yan Li}
\email{ly@ynao.ac.cn}

\author{Yan Li}
\affiliation{Yunnan Observatories, Chinese Academy of Sciences, Kunming 650216, China}
\affiliation{Key Laboratory for the Structure and Evolution of Celestial Objects, Chinese Academy of Sciences}
\affiliation{Center for Astronomical Mega-Science, Chinese Academy of Sciences, Beijing 100012, China}
\affiliation{University of Chinese Academy of Sciences, Beijing 100049, China}

\author{Xing-hao Chen}
\affiliation{Yunnan Observatories, Chinese Academy of Sciences, Kunming 650216, China}
\affiliation{Key Laboratory for the Structure and Evolution of Celestial Objects, Chinese Academy of Sciences}

\author{He-ran Xiong}
\affiliation{Yunnan Observatories, Chinese Academy of Sciences, Kunming 650216, China}
\affiliation{Key Laboratory for the Structure and Evolution of Celestial Objects, Chinese Academy of Sciences}

\author{Jun-jun Guo}
\affiliation{Yunnan Observatories, Chinese Academy of Sciences, Kunming 650216, China}
\affiliation{University of Chinese Academy of Sciences, Beijing 100049, China}

\author{Xue-fei Chen}
\affiliation{Yunnan Observatories, Chinese Academy of Sciences, Kunming 650216, China}
\affiliation{Key Laboratory for the Structure and Evolution of Celestial Objects, Chinese Academy of Sciences}
\affiliation{Center for Astronomical Mega-Science, Chinese Academy of Sciences, Beijing 100012, China}

\author{Zhan-wen Han}
\affiliation{Yunnan Observatories, Chinese Academy of Sciences, Kunming 650216, China}
\affiliation{Key Laboratory for the Structure and Evolution of Celestial Objects, Chinese Academy of Sciences}
\affiliation{Center for Astronomical Mega-Science, Chinese Academy of Sciences, Beijing 100012, China}
\affiliation{University of Chinese Academy of Sciences, Beijing 100049, China}


\begin{abstract}
Mixing in the convective core is quite uncertain in core helium burning stars. In order to explore the overshooting mixing beyond the convective core, we incorporated the $k$-$\omega$ proposed by Li (2012, 2017) into MESA, and investigated the overshooting mixing in evolution of sdB models. We found that the development of the convective core can be divided into three stages. When the radiative temperature gradient $\nabla_{\rm rad}$ monotonically decreases from the stellar center, the overshooting mixing presents a behaviour of exponential decay similar with Herwig (2000), and the overshooting distance is to make $\nabla_{\rm rad} \simeq \nabla_{\rm ad}$ at the boundary of the convective core, in agreement with the prediction of the self-driving mechanism Castellani, Giannone \& Renzini (1971a). When the radiative temperature gradient $\nabla_{\rm rad}$ shows a minimum value in the convective core, the convective core may be divided into two convection zones if the minimum value of $\nabla_{\rm rad}$ is smaller than the adiabatic temperature gradient $\nabla_{\rm ad}$. For the single-zone case, the overshooting mixing still shows an exponential decay behaviour, but the overshooting distance is much smaller than in the initial stage. For the double-zone case, the overshooting mixing is similar to that of the single-case beyond the convective core, while it almost stops on both sides of the above convective shell. Our overshooting mixing scheme is similar to the maximal overshoot scheme of Constantino et al. (2015). In the final stage, continuous injection of fresh helium into the convective core by the overshooting mixing happens, which is similar to the "core breathing pulses" (Sweigart \& Demarque 1973; Castellani et al. 1985). 
\end{abstract}


\section{Introduction}

Mixing in the convective core is a major uncertainty in the core helium burning models of low mass stars. The standard stellar evolution, which assumes complete mixing within the Schwarzschild boundary of the convective core, predicts a constant convective core mass (Paxton et al. 2013). As a result, a chemical discontinuity forms at the boundary of the convective core, as the increasing carbon and oxygen abundances produced by the helium burning process make the convective core more and more opaque than the outside helium-rich envelope. Schwarzschild \& H\"arm (1969) first noticed that this convective boundary cannot be stable . Any convective overshooting will mix carbon and oxygen inside the convective core into the above helium-rich zone, resulting in the increment of radiative temperature gradient and expansion of the convective core. There are many different treatments on the convective mixing beyond the Schwarzschild boundary of the convective core in the literature, leading to divergence of stellar evolution since the core helium burning (Paxton et al. 2013; Constantino et al. 2015; Schindler et al. 2015). 

In early studies, an overshooting scheme called by Castellani, Giannone \& Renzini (1971a) as the self-driving mechanism was often adopted, which allows the convective core to expand continually until the radiative temperature gradient $\nabla_{\rm rad}$ is equal to the adiabatic temperature gradient $\nabla_{\rm ad}$ at the boundary of the convective core. When the central helium abundance decreases to below a critical value, a minimum value of $\nabla_{\rm rad}$ will be present in the convective core. Therefore, further convective overshooting will result in lowing of $\nabla_{\rm rad}$ and finally make this minimum value equal to the value of $\nabla_{\rm ad}$. Schwarzschild \& Harm (1969) first pointed out that a partially mixed region can develop beyond the convective core, and then introduced the concept of semiconvection. In this case, the complete mixing is assumed to happen within the point of $\nabla_{\rm rad} = \nabla_{\rm ad}$, and a partial mixing called by Castellani, Giannone \& Renzini (1971b) as the induced semi-convection is introduced from the boundary of the complete mixing to the edge of the convective core in order that the condition $\nabla_{\rm rad} = \nabla_{\rm ad}$ is satisfied everywhere in this region (e.g.  Simpson 1971; Robertson \& Faulkner 1972; Faulkner \& Cannon 1973).

However, recent numerical simulations show (Freytag et al. 1996) that the overshooting mixing decays exponentially outside the convective boundary, and the overshooting mixing beyond the boundary of the convective core is often calculated using an exponentially decaying diffusion coefficient (Herwig 2000). This scheme results in similar effect of expansion of the convective core in the early stage of the core helium burning as the self-driving mechanism. However, the minimum value of $\nabla_{\rm rad}$ may be below the local value of $\nabla_{\rm ad}$ due to the overshooting mixing and the convective core is often divided into two, when the central helium abundance drops below a critical value (Paxton et al. 2013; Schindler et al. 2015). Constantino et al. (2015) assumed that the overshooting mixing may be suspended when the difference of $\nabla_{\rm rad}$ and $\nabla_{\rm ad}$ is less than 0.002, in order to maintain the integrity of the convection core.

Recently Li (2012, 2017) developed a $k$-$\omega$ model to describe the overshooting mixing beyond the convective core. The $k$-$\omega$ model is fully based on turbulent convection model, which not only includes carefully the effects of buoyancy and the shear of convective rolling cells in the dynamic equation of the turbulent kinetic energy $k$, but also establish a dynamic equation for the turbulence frequency $\omega$. By solving the above $k$-$\omega$ equations numerically, the turbulent diffusion coefficient can finally be obtained not only in the convectively unstable core but also in the overshooting region beyond.

Subdwarf B (sdB) stars are generally believed to be core helium burning stars with extremely thin hydrogen envelope ($<0.02M_{\odot}$, Heber 2009, 2016). Most of them are found in binary systems (Maxted et al. 2001; Napiwotzki et al. 2004; Copperwheat et al. 2011), and their formation channels in the binary evolution scenario have been well established (Han et al. 2002, 2003). Due to their extremely thin hydrogen envelope, the hydrogen burning almost quenches at the base of the envelope, leaving the core helium burning as the only nuclear burning source. This is a good example for us to explore the overshooting mixing beyond the convective core. Some sdB stars are found to be g-mode pulsators (Kilkenny et al. 1997; Green et al. 2003; Schuh et al. 2006). As g modes cannot propagate in the convective zone, their periods are quite sensitive to the size of the convective core in sdB stars. Recently, van Grootel et al. (2010a, 2010b) and Charpinet et al. (2011) explored from asteroseismology the masses of the convective cores for three sdB stars (KPD 0629-0016, KPD 1943+4058, and KIC02697388), and found that their derived convective core mass ($0.22\sim 0.28M_{\odot}$) are generally much larger than what the stellar models without overshooting predict (about $0.1M_{\odot}$). 

In order to study the overshooting of the convective core and its effect on the evolution of the sdB stars, we applied the $k$-$\omega$ model in calculations of the evolution of an sdB model. We briefly describe the $k$-$\omega$ model we have used in Section 2, and introduce the input physics of our stellar models in Section 3. Our results of calculations are described in details in Section 4. In Section 5, we summarize our main conclusions.

\section{ Equations of the $\MakeLowercase{k}$-$\omega$ model for stellar convection }

The $k$-$\omega$ model proposed by Li (2012, 2017) is a two-equation model of turbulence in the stellar convection zone. It is fully based on the moment equations of fluid hydrodynamic. In the equation of turbulent kinetic energy $k$, generations of turbulence due to buoyancy and shear of convective rolling cells are carefully approximated with proper turbulence models; while in the equation of turbulence frequency $\omega$, a macro-length model is reasonably introduced. With the gradient type hypothesis for the transport effect of turbulence, the $k$-$\omega$ model can be applied to not only convection zones but also overshooting regions. It is worth noting that its local solution is identical with the classical mixing-length theory (Cox \& Giuli 1968).

According to Li (2012, 2017), we describe the stellar convection with the equation of turbulent kinetic energy $k$:
\begin{equation}\label{2c1}
\frac{\partial k}{\partial t}-\frac{1}{{{r}^{2}}}\frac{\partial }{\partial r}\left( {{r}^{2}}{{\nu }_{t}}\frac{\partial k}{\partial r} \right)=P+G-k\omega  ,
\end{equation}
and the equation of turbulence frequency $\omega$:
\begin{equation}\label{2c2}
\frac{\partial \omega }{\partial t}-\frac{1}{{{r}^{2}}}\frac{\partial }{\partial r}\left( {{r}^{2}}\frac{{{\nu }_{t}}}{{{\sigma }_{\omega }}}\frac{\partial \omega }{\partial r} \right)=\frac{c_{L}^{2}k}{{{L}^{2}}}-{{\omega }^{2}}.
\end{equation}
In Equations (\ref{2c1}) and (\ref{2c2}), the turbulent diffusivity $\nu_{t}$ is approximated by:
\begin{equation}\label{2c3}
{{\nu }_{t}}={{c}_{\mu }}\frac{k}{\omega }.
\end{equation}
In Equation (\ref{2c1}), $P$ and $G$ represent respectively the shear and buoyancy production rate of the turbulent kinetic energy which are discussed in detail in Li (2012, 2017). In Equation (\ref{2c2}), $L$ represents the macro-length of turbulence, which is equivalent to the mixing-length in the standard mixing-length theory. According to common choices (Pope 2000), the model parameters $ c_{\mu} = 0.09 $ and ${c}_{L}=c_{\mu }^{3/4}$, and the model parameter ${\sigma }_{\omega }=1.5$. 

The macro-length of turbulence $L$ is usually assumed to be proportional to the local pressure scale height $H_P$ for the convective envelope, because the thickness of the stellar convective envelope is usually much longer than the local pressure scale height. For the convective core, however, its thickness is usually smaller than the local pressure scale height. Accordingly, the macro-length of turbulence $L$ will be restricted by the actual size of the convection core. We therefore adopt the model of Li (2017) for the macro-length of turbulence in the convective core:
\begin{equation}\label{3c32}
L={{c}_{L}}{\alpha }'{{R}_{\rm cc}},
\end{equation}
where $R_{\rm cc}$ is the radius of the convective core, and ${\alpha }'$ is an adjustable parameter.

The evolution of element in the stellar interior can be approximately treated by a diffusion equation:
\begin{equation}\label{3c10}
\frac{\partial {{X}_{i}}}{\partial t}=\frac{\partial }{\partial m}\left[ {{\left( 4\pi \rho {{r}^{2}} \right)}^{2}}\left( {{D}_{\rm mix}}+{{D}_{t}} \right)\frac{\partial {{X}_{i}}}{\partial m} \right]+{{d}_{i}},
\end{equation}
where $X_i$ is the mass fraction and $d_i$ is the generation rate of element “$i$”. $D_{\rm mix}$ represents the diffusion coefficient due to any mixing process, such as element diffusion, meridional circulation, etc, while $D_t$ is due to the convective and/or overshooting mixing that is given by Li (2017):
\begin{equation}\label{3c15}
{{D}_{t}}=\frac{{{c}_{X}}}{1+\frac{\lambda \omega }{\rho {{c}_{P}}k}+{{c}_{t}}{{c}_{\theta }}{{\omega }^{-2}}{{N}^{2}}}\frac{k}{\omega },
\end{equation}
where $\rho$ is the density, $P$ the total pressure, $c_P$ the specific heat at constant pressure. $c_t$ and $c_{\theta}$ are two model parameters, and their values are given by turbulence models (Hossain \& Rodi 1982): $c_t=0.1$ and $c_{\theta}=0.5$. In addition, $c_X$ is an adjustable parameter. In Equation (\ref{3c15}), the buoyancy frequency $N$ is defined in a chemically homogeneous region as:
\begin{equation}\label{3c12}
{{N}^{2}}=-\frac{\beta gT}{{{H}_{P}}}\left( \nabla -{{\nabla }_{\rm ad}} \right),
\end{equation}
where $T$ is the temperature, $g$ the gravity acceleration, $\nabla$ and $\nabla_{\rm ad}$ are respectively the actual and adiabatic temperature gradient, and the thermodynamic coefficient $\beta$ is defined by
\begin{equation}\label{3c14}
\beta =-\frac{1}{\rho }{{\left( \frac{\partial \rho }{\partial T} \right)}_{P}}.
\end{equation}
In addition, the radiation diffusivity $\lambda$ is defined by
\begin{equation}\label{3c16}
\lambda =\frac{16\sigma {{T}^{3}}}{3\rho \kappa },
\end{equation}
where $\kappa$ is the Rosseland mean opacity, and $\sigma$ the Stefan-Boltzmann constant. It should be noted that we neglect the composition gradient in $N^2$ for this study.

\section{ Input physics and numerical schemes }

Our stellar models are computed by the Modules of Experiments in Stellar Astrophysics (MESA), which was developed by Paxton et al. (2011, 2013). We follow the works of Schindler et al. (2015) and Xiong et al. (2017) to calculate the evolution of sdB models, using version 6596 of MESA with the default parameters. The detailed input physics is described as follows. We choose the initial metallicity of our sdB models to be $Z=0.02$, and the initial helium abundance to be $Y=0.28$. In order to evolve the stellar models to the helium burning stage, we use the opacity of "OPAL\_Type2" and the nuclear burning networks of "o18\_and\_ne22". We use the $k$-$\omega$ model to treat convection, whose parameters are summarized in Table 1. It should be noted that the values of three adjustable parameters ($c_X$, $c_h$, and $\alpha'$) are according to solar and stellar calibrations (Li 2012, 2017). Mixing due to atomic diffusion is not included in our stellar models, though it is believed to be significant in sdB stars to reproduce their pulsation properties (Charpinet et al. 1997; Chayer et al. 2004; Fontaine et al. 2006; Michaud et al. 2007, 2008; Hu et al. 2009, 2011).

\begin{deluxetable}{ccccccc}
\tablewidth{0pt}
\tablecaption{Parameters of the $k$-$\omega$ model}
\tablehead{
\colhead{$c_{\mu}$} &    \colhead{$\sigma_{\omega} $} & 
\colhead{$c_{\theta}$} & \colhead{$c_t$} & 
\colhead{$c_h$} &        \colhead{$c_X$} &          
\colhead{${\alpha}'$}  
}
\startdata
0.09 & 1.5 & 0.5 & 0.1 & 2.344 & 0.01 & 0.06 \\
\enddata
\end{deluxetable}

The convective and/or overshooting mixing is treated by use of the $k$-$\omega$ model in the stellar models. We implement the $k$-$\omega$ model into MESA through the package "run\_star\_extras.f" as follows. When a stellar model is obtained by solving the equations of the stellar structure, we solve Equations (\ref{2c1}) and (\ref{2c2}) in the package "run\_star\_extras.f" to obtain $D_t$. Then we apply $D_t$ to continue calculations of the chemical evolution with MESA.

In order to obtain an initial sdB model with a specified mass, we calculate the evolution of a $4.5M_{\odot}$ star first, using the package  "7M\_prems\_to\_AGB". At this stage we adopt the standard mixing-length theory to treat convection, and choose the mixing-length parameter $\alpha=1.8$. We calculate the evolution of the star from the pre-main-sequence to the red giant phase, and stop at a model with its helium core mass to be $0.45M_{\odot}$. Then we assume that the star undergoes a mass transfer to its companion star, and approximate this process by artificially removing the stellar envelope with a high mass-loss rate ($5\times {{10}^{-5}}{{M}_{\odot }}{\rm yr}^{-1}$). We finally stop this evolutionary calculations when the condition "envelope\_mass\_limit=0.001" is satisfied. As a result, the obtained initial sdB model has a mass of about $0.455M_{\odot}$. It is well known that the chemical profile of an sdB star is sensitive to its early evolution, especially whether it undergoes the helium flashes or not (Dorman et al. 1993; Hu et al. 2009, 2011; {\O}stensen et al. 2012). In the present work, we focus on the evolution of the convective core. Therefore, small features left by previous evolution on the envelope of our initial sdB model will have little effect on our main results.

Afterwards, we start calculations of the evolution of the initial sdB model obtained above, using the $k$-$\omega$ model to treat convection and/or overshooting mixing in the stellar core. During this stage, we use the package "1M\_pre\_ms\_to\_AGB", with the same parameters mentioned above. Our calculations of stellar evolution stop when helium is exhausted at the stellar center.

\section{ The properties of stellar structure during the sdB evolution }

The development of convection in the stellar interior is shown in Figure 1a, which displays the boundary of the convective core as a function of time. The same plot is also given in Figure 1c for a stellar model with identical parameters as in Figure 1a, but with the overshooting mixing scheme of Herwig (2000). It can be noticed that our model is most similar to a model with an overshooting mixing of $f_{\rm ov}=0.004$. Throughout the whole evolution, there are basically three main stages in the development of convection in our sdB models.

\begin{figure}
\plotone{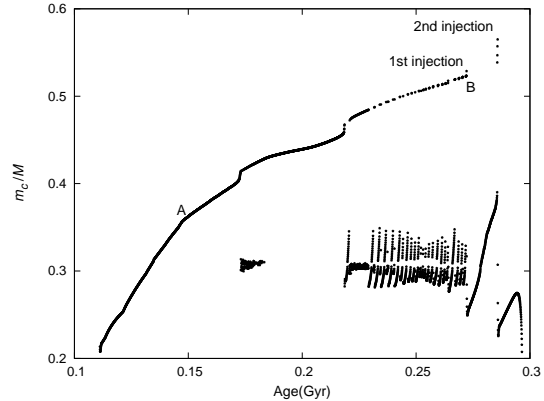}
\figurenum{1a}
\caption{Development of the convective core during the helium burning stage. The initial stage is before point A, the middle stage is between point A and B, and the final stage is after point B. During the middle stage, convection is frequently divided into two parts (a convective core and a convective shell). Therefore the size of the convective core greatly decreases, seen as vertical points much below the upper line. Then the convective core expands due to overshooting, until it touches the lower edge of the upper convective shell and merge with it to recover its size as before. The first and second injection of helium are also indicated. }
\end{figure}

\begin{figure}
\plotone{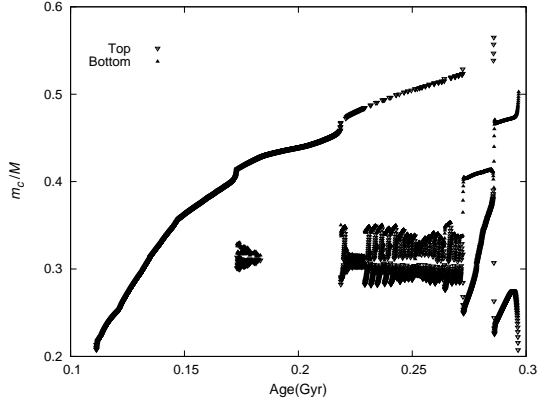}
\figurenum{1b}
\caption{Same as Figure 1a, but with triangle to represent the bottom and inverse triangle to represent the top of the convection zones. The radiative equilibrium zone between the central convective core and the upper convective shell is clearly presented by the middle shaded area.}
\end{figure}

\begin{figure}
\plotone{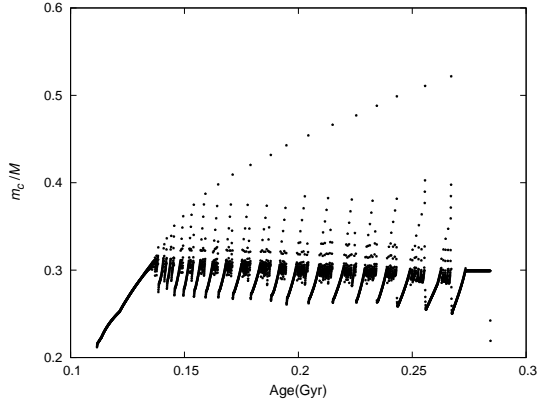}
\figurenum{1c}
\caption{Same as Figure 1a, but for stellar models with the overshoot mixing scheme of Herwig (2000). }
\end{figure}

\subsection{ The initial stage }

In the initial stage, the radiative temperature gradient monotonically decreases outwardly from the center, which defines a definite boundary for the convective core. As seen in Figure 1a, the mass of the convective core increases monotonically during this stage (before point A).

\begin{figure}
\plotone{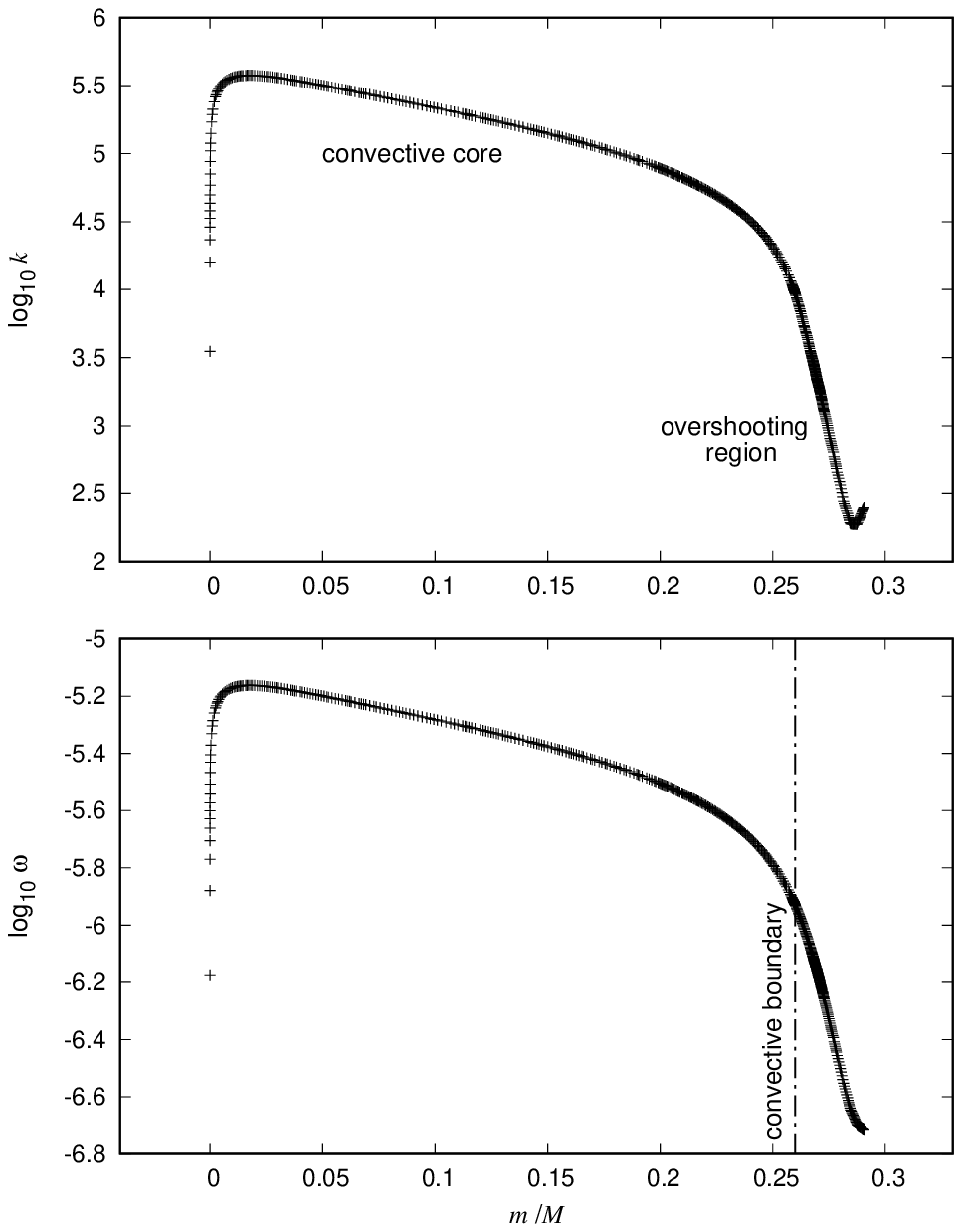}
\figurenum{2}
\caption{Distributions of the turbulent kinetic energy (upper panel) and the turbulence frequency (lower panel) in the convective core and the overshooting region for an sdB model in the initial stage. }
\end{figure}

\begin{figure}
\plotone{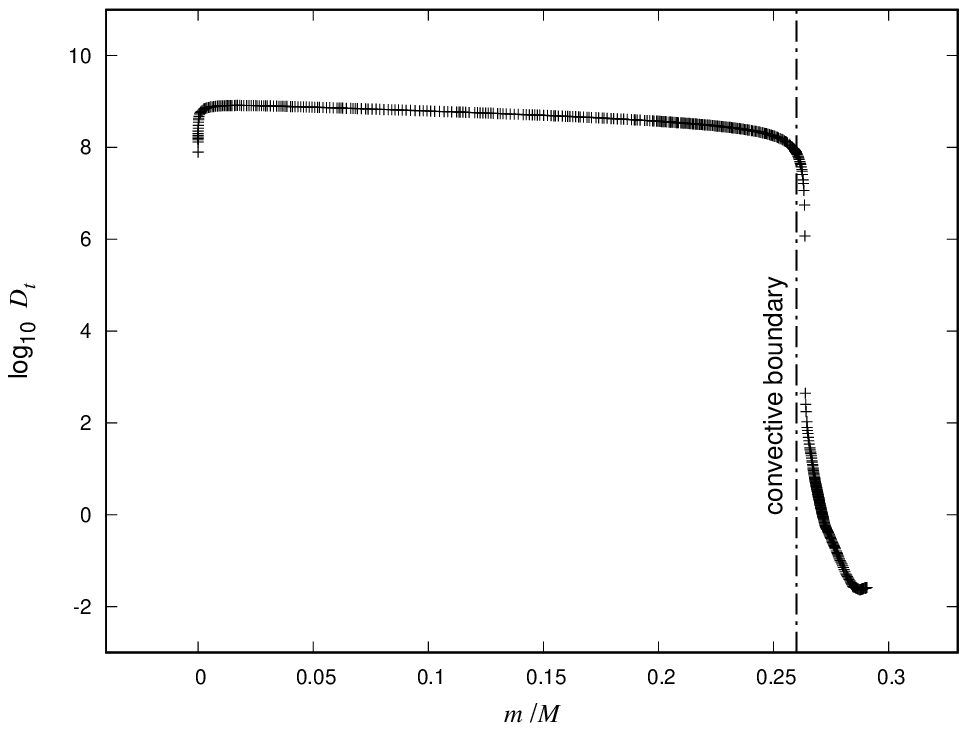}
\figurenum{3}
\caption{Distribution of the turbulent diffusivity in the convective core and the overshooting region for an sdB model in the initial stage.}
\end{figure}

The profiles of turbulent kinetic energy and turbulence frequency are shown in Figure 2. It can be seen that the turbulent kinetic energy is around $10^5$ in the convective core, which is in agreement with the prediction of the MLT. The effect of turbulent diffusion process is significant in two regions: near the stellar center and beyond the Schwarzschild boundary of the convective core. The turbulent kinetic energy decreases rapidly near the center, because the local pressure scale height will be infinite when near to the center while the turbulent kinetic energy is inversely proportional to the local pressure scale height. Here the turbulent diffusion effect is significant, which makes the turbulent kinetic energy much larger than the MLT predicted value. But much more importantly, the turbulent diffusion transports a significant amount of turbulent kinetic energy outside the convective core, leading to formation of an overshooting region. It should be noticed in Figure 2 that the $k$-$\omega$ model predicts an exponential decay of both the turbulent kinetic energy and the turbulence frequency in the overshooting region. The corresponding distribution of turbulent diffusivity is shown in Figure 3. It can be seen that $D_t$ is about $10^9$ in the convective core, which guarantees a complete mixing in the convective core. In the overshooting region, however, $D_t$ drops suddenly to $10^2$ and then continues to decay rapidly, resulting in a partial mixing in the overshooting region.

\begin{figure}
\plotone{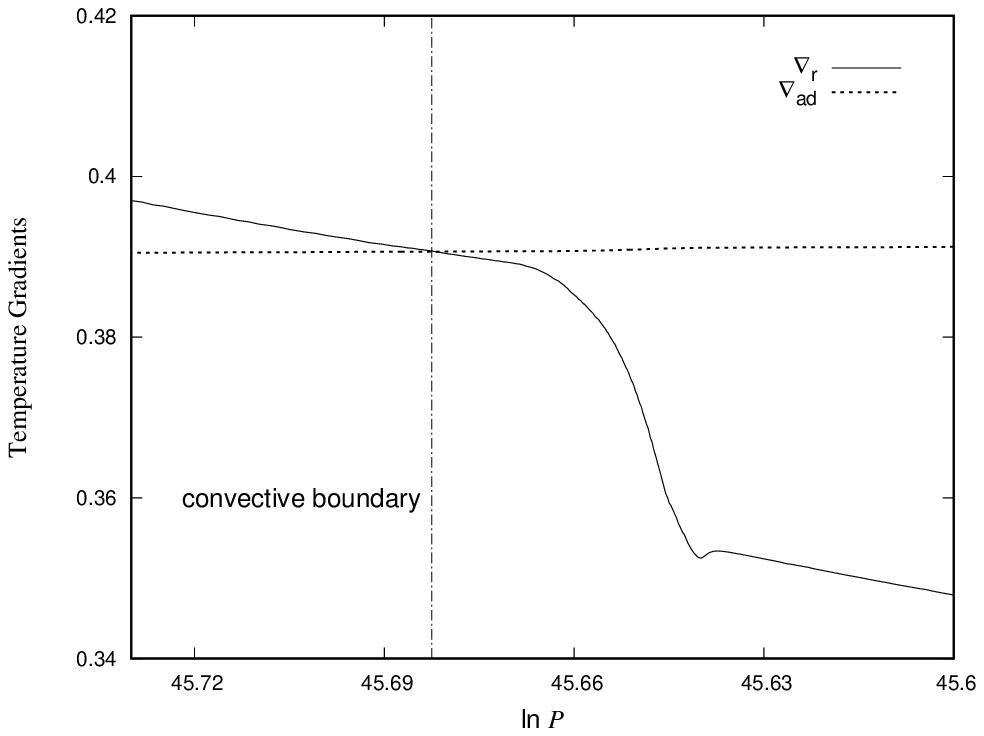}
\figurenum{4}
\caption{Distributions of both radiative and adiabatic temperature gradients for an sdB model in the initial stage. The vertical line indicates the Schwarzschild boundary of the convective core.}
\end{figure}

\begin{figure}
\plotone{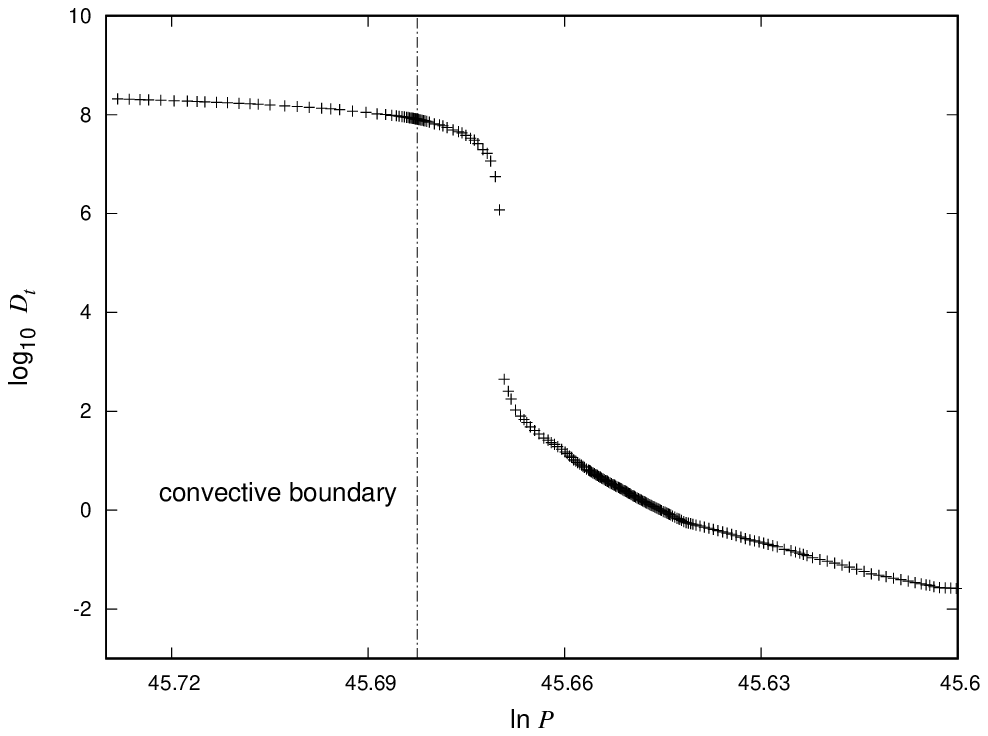}
\figurenum{5}
\caption{Distribution of the turbulent diffusivity near the boundary of the convective core for an sdB model in the initial stage. The vertical line indicates the Schwarzschild boundary of the convective core. }
\end{figure}

\begin{figure}
\plotone{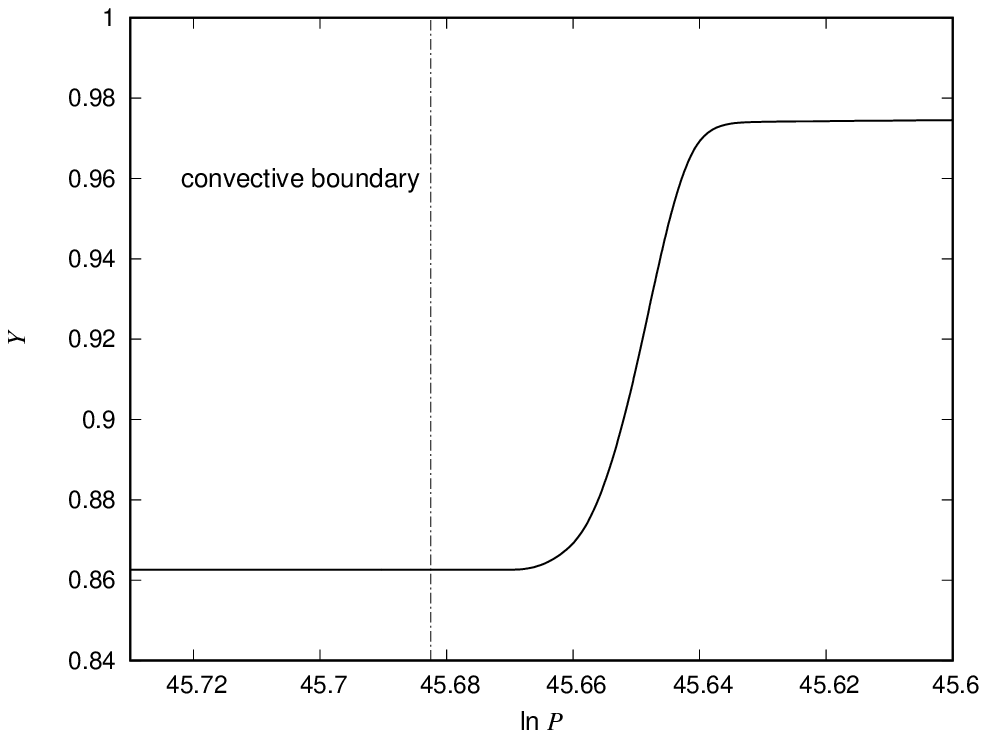}
\figurenum{6}
\caption{Distribution of the helium abundance near the boundary of the convective core for an sdB model in the initial stage. }
\end{figure}

In order to show the detailed structure of the overshooting region, we show profiles of both radiative and adiabatic temperature gradients in Figure 4. The Schwarzschild boundary of the convective core is indicated by a vertical dashed line in Figure 4, where the radiative temperature gradient equals the adiabatic temperature gradient. The corresponding profile of turbulent diffusivity is then shown in Figure 5. We show also the Schwarzschild boundary of the convective core by the same vertical dashed line. It is worth to note in Figure 5 that the turbulent diffusivity remains at a high value not only within the Schwarzschild boundary but also a little bit outwards, resulting in an extra completely mixing region of about $0.012 H_P$. Then the turbulent diffusivity decays roughly exponentially, extending further for about $0.05 H_P$ before it becomes below about 0.1. It can be seen in Figure 4 that the overshooting mixing scheme we have adopted is similar to the classical overshooting mixing scheme, which allows the convective core to mix further until the radiative temperature gradient is equal to the adiabatic one at the boundary of the resulted fully mixed core (Castellani, Giannone \& Renzini 1971a). The corresponding helium abundance profile is shown in Figure 6. It can be clearly seen that the completely mixing region extends beyond the Schwarzschild boundary of the convective core, which confirms the result of the turbulent diffusivity. Then the helium abundance rises from the central value smoothly to the envelope value.

In summary, the $k$-$\omega$ model predicts a distinctive structure of the overshooting region beyond the convective core, i.e., an extra completely mixing part plus a diffusively mixing part. Such a mixing scheme is similar to the widely adopted overshooting mixing scheme proposed by Herwig et al. (2000). With the same size of the complete mixing region, however, our scheme can usually provide a much wider partially mixing region than theirs. An extra mixing is important in sdB models, for it is the only way to increase the size of the convective core during the sdB phase to fulfil the asteroseismological requirements. A smooth transition in the chemical profile is also essential for the Brunt-V\"ais\"al\"a frequency not to be undefined in a chemical gradient region.

\begin{figure}
\plotone{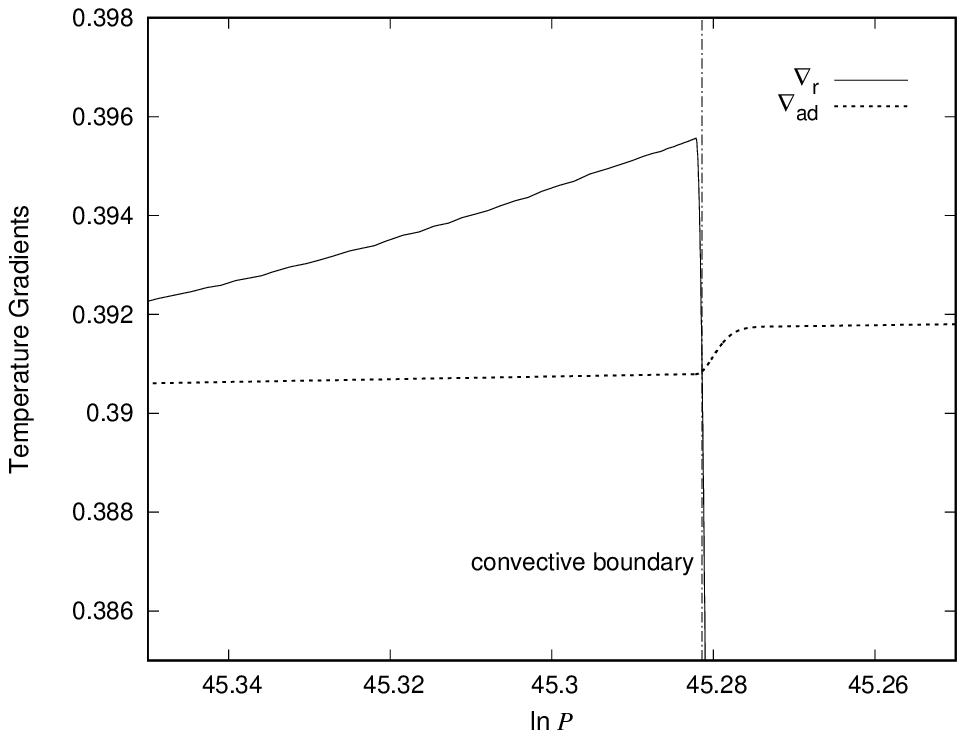}
\figurenum{7}
\caption{Distributions of both radiative and adiabatic temperature gradients for an sdB model in the middle stage. }
\end{figure}

\begin{figure}
\plotone{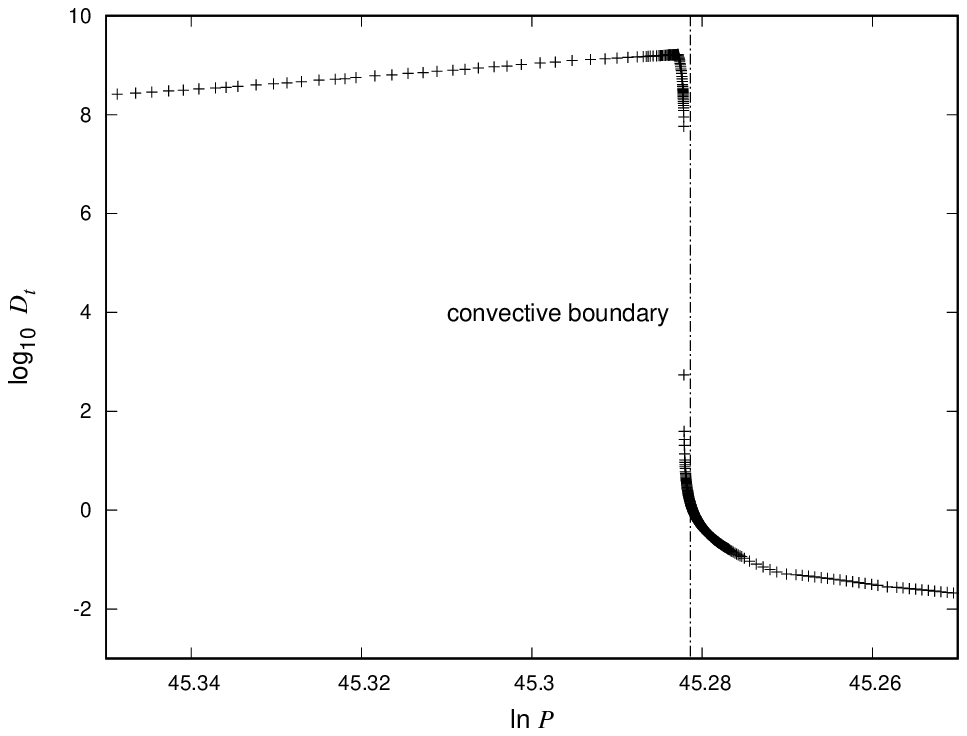}
\figurenum{8}
\caption{Distribution of the turbulent diffusivity in the convective core and the overshooting region for an sdB model in the middle stage. }
\end{figure}

\begin{figure}
\plotone{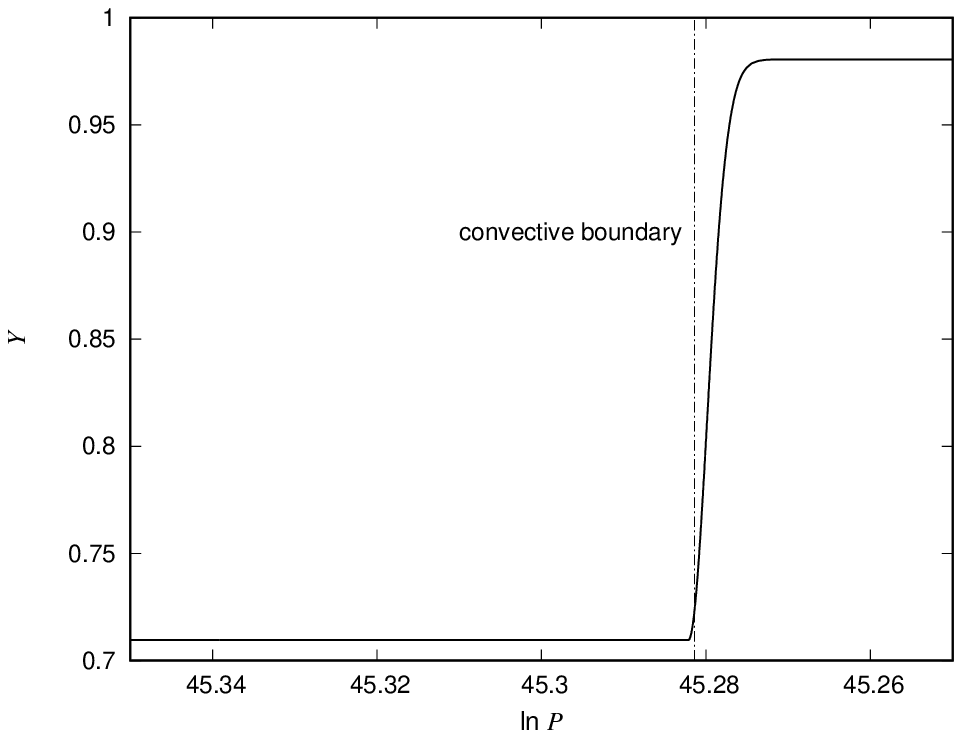}
\figurenum{9}
\caption{Distribution of the helium abundance near the boundary of the convective core for an sdB model in the middle stage. }
\end{figure}

\subsection{ The middle stage }

In this stage, the radiative temperature gradient firstly decreases but then increases in the stellar core, leading to the appearance of a minimum value (from point A to point B in Figure 1a). Whether this minimum value is greater than the value of the adiabatic temperature gradient is critical for the development of convection in the stellar core. When it is greater than the adiabatic one, the boundary of the convective core can still be well defined. We refer to this as the single-zone case. As shown in Figure 1a, the mass of the convective core keeps growing due to the overshoot mixing beyond the convective boundary. Once the minimum value of the radiative temperature gradient becomes smaller than the value of the adiabatic temperature gradient, convection is divided into two zones in the stellar core: a central convective core plus an outer convective shell. We refer to this as the double-zone case. It can be noticed in Figure 1b that the mass of the convective core drops repeatedly to a low value, indicating the appearance of a middle radiative shell above the convective core.

The profiles of both radiative and adiabatic temperature gradients are shown for a model of the single-zone case in Figure 7, and the distribution of the turbulent diffusivity is shown as well for the same model in Figure 8. It can be noticed in Figure 7 that the radiative temperature gradient suddenly drops near the boundary of the convective core, which is in contrast to declined gradually in the initial stage. As shown in Figure 8, the turbulent diffusivity also drops dramatically near the boundary of the convective core, preventing the presence of an extra completely mixing overshooting region. It should also be noticed that this does happen inside the Schwarzschild boundary of the convective core. The profile of helium abundance is shown in Figure 9. It can be clearly seen that the fully mixing region only extends to the boundary of the convective core, and the partially mixing region occupies a width of only about $0.006H_P$. Compared with models in the initial stage, the overshooting mixing is significantly restricted for models of the single-zone case in the middle stage.

\begin{figure}
\plotone{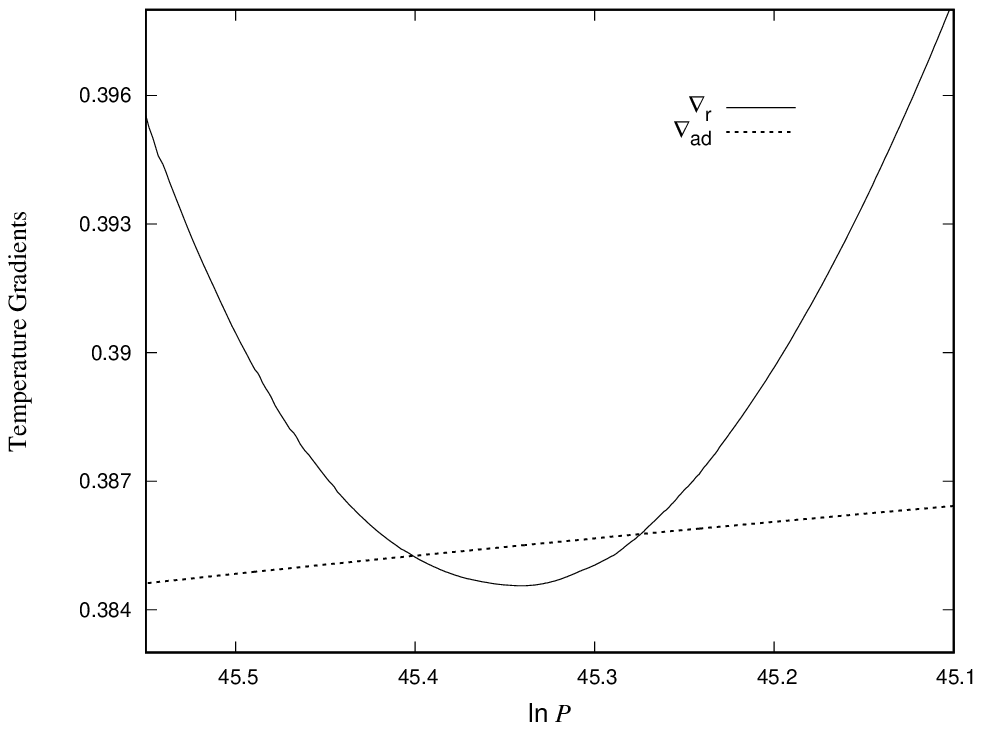}
\figurenum{10}
\caption{Distributions of both radiative and adiabatic temperature gradients for an sdB model in the middle stage with two separated convection zones. }
\end{figure}

\begin{figure}
\plotone{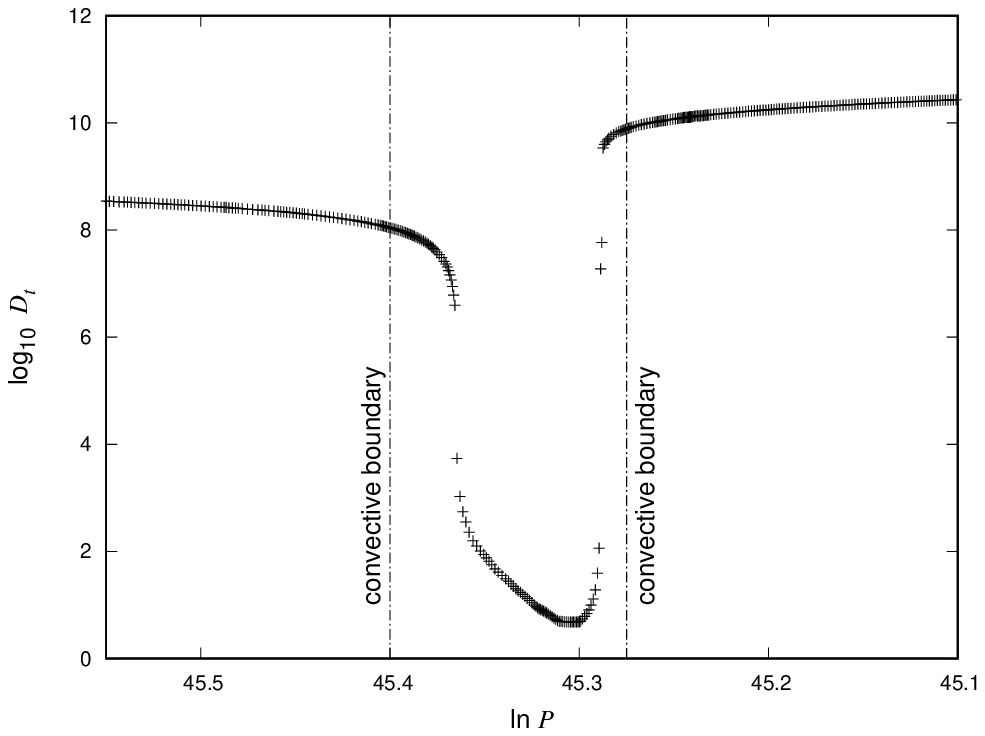}
\figurenum{11}
\caption{Distributions of the turbulent diffusivity for an sdB model in the middle stage with two separated convection zones. }
\end{figure}

\begin{figure}
\plotone{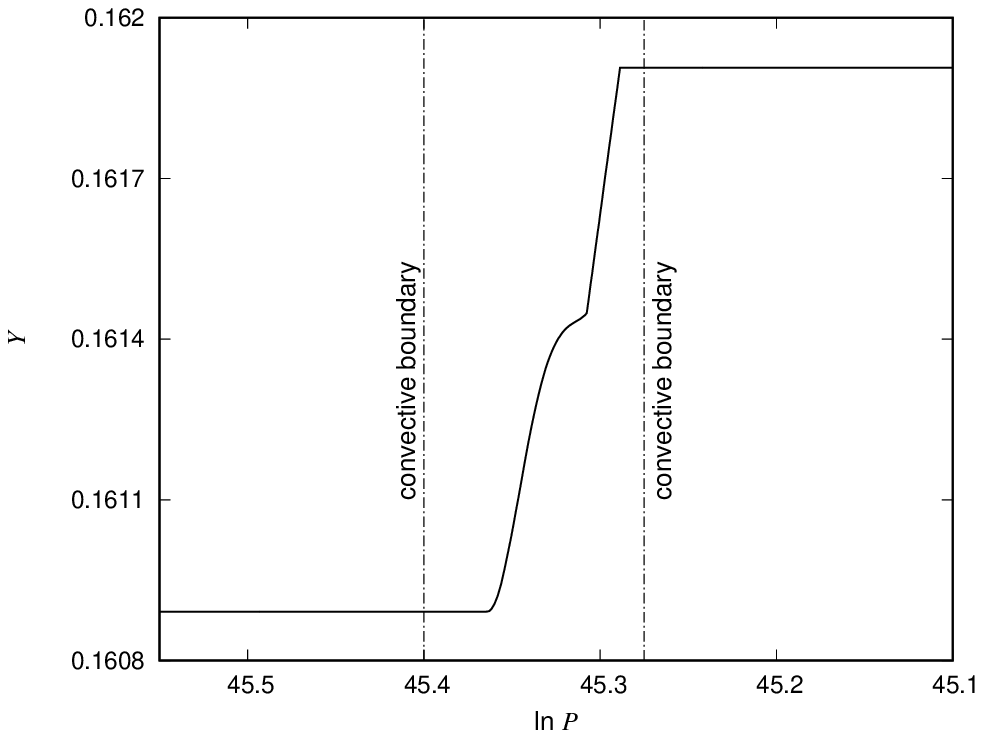}
\figurenum{12}
\caption{Distributions of the helium abundance for an sdB model in the middle stage with two separated convection zones. }
\end{figure}

Since the opacity of free-free transition is quite sensitive to the carbon abundance, the radiative temperature gradient will be lowered if an enough amount of fresh helium is injected by the overshooting mixing into the carbon-rich convective core. Once the radiative temperature gradient is smaller than the adiabatic temperature gradient, the convective core will be divided into two zones. One model of this double-zone case is illustrated in Figure 10, in which a radiative zone with a thickness of about $0.12H_P$ is in the middle of two convection regions. The corresponding turbulent diffusivity is shown in Figure 11. It can be seen that $D_t$ drops dramatically in the radiative zone, but its behaviour is different on the two opposite ends. On the left side, $D_t$ remains at a high value beyond for about $0.03H_P$ from the boundary of the convective core, and then decays exponentially further outwards. This is quite similar with the case in the initial stage. But on the right side, $D_t$ drops suddenly and soon meet the tail of the decay from the left side, resulting in almost no more mixing below the boundary of the convective shell. The profile of helium abundance is shown in Figure 12, representing clearly the overshooting mixing beyond the boundary of the convective core.

It is evident that the effect of the helium injection mentioned above is more efficient when the convective core is more carbon abundant. As shown in Figures 10 and 11, the separation of the two convection zones is usually much larger than the overshooting distance beyond the convective core. Then the convective core grows step by step due to overshooting, as shown in Figure 1a. On the other hand, the upper convective shell keeps almost unchanged. Thus the convective core will finally touch the boundary of the upper convective shell, and then the two convection zones will merge again into one convective core. As shown in Figure 1b, this process repeats for many times when the central helium abundance is below 0.35.

\subsection{ The final stage }

When the central helium abundance is below about 0.1, most of helium nuclei are captured by carbon to produce oxygen, rather than to produce carbon through $3\alpha$ reaction. The injection of fresh helium into the convective core by the overshooting mixing can decrease the radiative temperature gradient, but the production of oxygen can increase it because oxygen is much more opaque than carbon. Then we may expect that the convective core will continue to expand until enough fresh helium is injected into it to finally make the radiative temperature gradient begin to decrease. This is the final stage of our sdB evolution (after point B in Figure 1a).

It can be seen in Figure 1a that the continuous expansion of the convective core takes place for two times in our sdB evolution. After the first time of the continuous expansion, the boundary of the convective core reduces much more than in the models of the middle case. Later, the convective core re-expands due to overshooting mixing beyond its boundary. When its mass increases to about 0.4 of the total stellar mass, it merges with the upper convective shell and then continues to expand. When enough fresh helium is injected into it, the mass enclosed by the convective core reduces even more significantly. Afterwards, it consumes up all of helium before its boundary reaches again the upper convective shell.

\subsection{ The evolution of the sdB models }

\begin{figure}
\plotone{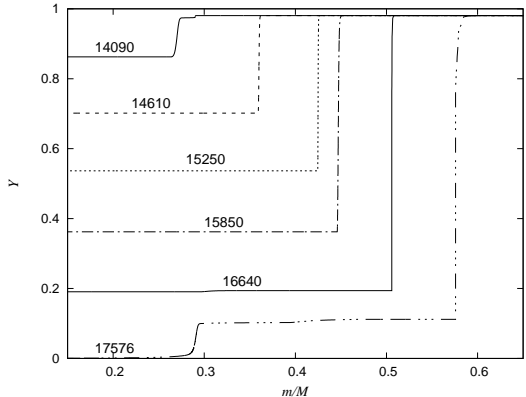}
\figurenum{13a}
\caption{Profiles of the helium abundance during the core helium burning stage for stellar models with the overshoot mixing scheme of the $k$-$\omega$ model. Model numbers are also indicated for each profile. }
\end{figure}

\begin{figure}
\plotone{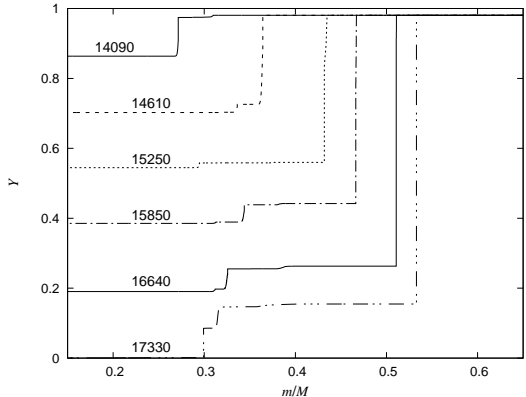}
\figurenum{13b}
\caption{Same as Figure 13a, but for stellar models with the overshoot mixing scheme of Herwig (2000). }
\end{figure}

The evolution of the helium profile is shown in Figure 13a. It can be seen that the helium profiles of our sdB models are quite similar with those of Constantino et al. (2015) with the maximal overshoot scheme. This is because our core mixing scheme is similar to the maximal overshoot scheme they have adopted. In their core mixing scheme, the overshooting distance is chosen such large that the minimum value of the radiative temperature gradient is almost equal to (but still larger than) the adiabatic temperature gradient. In our overshooting mixing scheme, the minimum value of the radiative temperature gradient keeps also close to the adiabatic temperature gradient as shown in Figure 10, but it can often be smaller than the adiabatic one to stop the overshooting mixing of helium into the stellar core. This property is similar with the maximal overshoot scheme of Constantino et al. (2015), who stop the core mixing when the difference of the radiative temperature gradient and the adiabatic temperature gradient falls below 0.002.
For comparisons, we show also in Figure 13b the result of the stellar model with the overshooting mixing scheme of Herwig (2000). It can be noticed that more inflections are present in the helium profile, especially in the late helium burning stage.

The evolutionary tracks of our sdB models are shown in Figure 14. It can be seen that overall property of the evolutionary track is similar with those of Xiong et al. (2017). However, we notice that there is one loop formed in the late stage of the evolution, similar to the so called "core breathing pulses" (Sweigart \& Demarque 1973; Castellani et al. 1985).

The chemical profiles of the last model with the exhaustion of helium at the stellar center is shown in Figure 15a. It can be seen that the C/O core occupies about innermost 28\% of the total stellar mass, while the helium envelope occupies the outermost 40\% of the total stellar mass. In addition, there is a He/C/O zone of about 32\% of the total stellar mass in between, which is a result of the second core breathing pulse as seen in Figure 1a. We show also in Figure 15b the result of the stellar model with the overshooting mixing scheme of Herwig (2000) for comparisons. It should be noted that our C/O core is a little larger than the model with Herwig (2000) overshooting mixing scheme, because we include the helium injections in the final stage of the helium burning phase.

\begin{figure}
\plotone{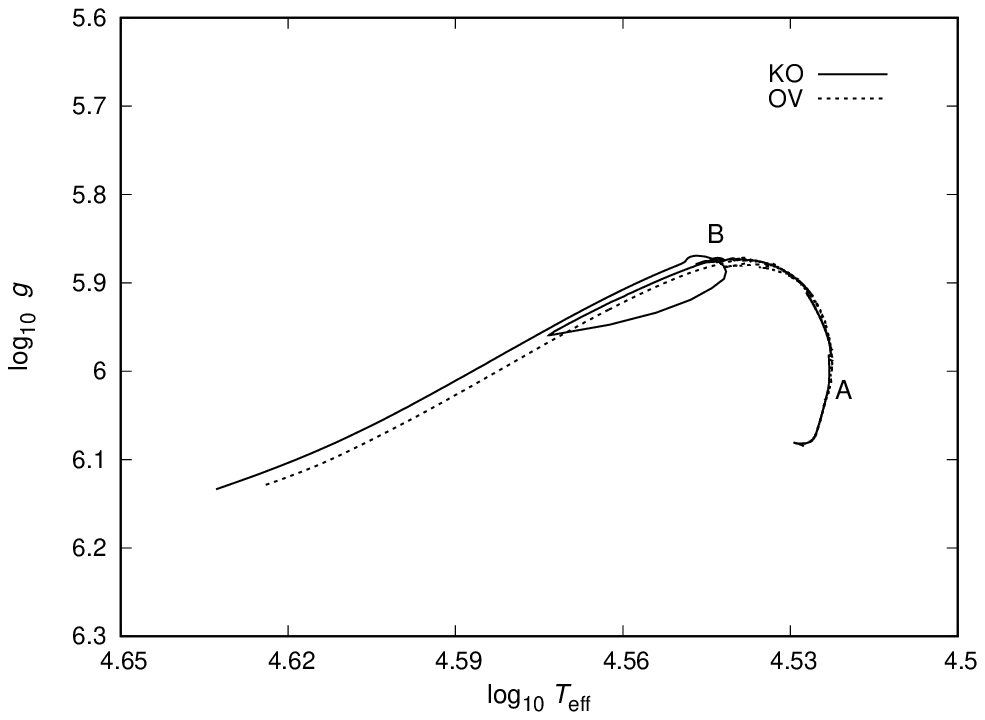}
\figurenum{14}
\caption{Evolutionary tracks of the considered sdB models. Solid line is for stellar models with the overshooting mixing scheme of the $k$-$\omega$ model, while dotted line for stellar models with the overshooting mixing scheme of Herwig (2000). Point A and B are the same as in Figure 1a.}
\end{figure}

\begin{figure}
\plotone{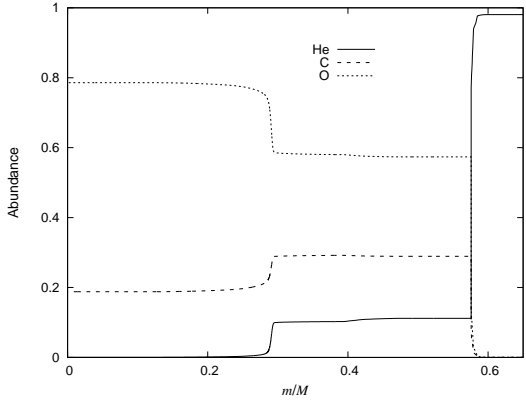}
\figurenum{15a}
\caption{Chemical profiles of the last sdB model with helium exhaustion at its center.}
\end{figure}

\begin{figure}
\plotone{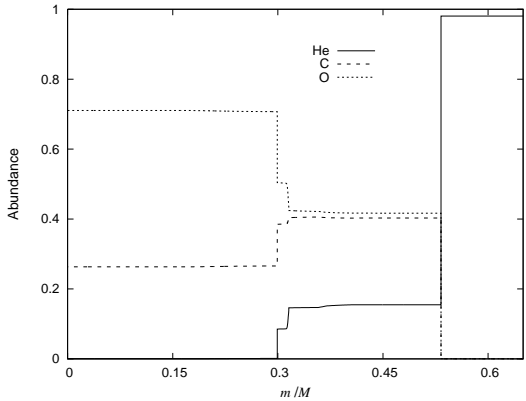}
\figurenum{15b}
\caption{Same as Figure 15a, but for stellar models with the overshoot mixing scheme of Herwig (2000). }
\end{figure}

\section{ Conclusions and Discussions }

Mixing in the convective core is quite uncertain in core helium burning stars. Many approaches, such as convective overshooting, semiconvection, etc. have been adopted in stellar models, leading to different stellar structure and various scenarios of stellar evolution. Li (2012, 2017) proposed the $k$-$\omega$ model for stellar convection, which can be applied to both convectively unstable zone and overshooting regions. We incorporated it into MESA, and investigated the overshooting mixing in evolution of sdB models. We found that the development of the convective core can be divided into three stages, and in each stage the convective overshooting show distinctive characteristics. 

In the initial stage, the radiative temperature gradient $\nabla_{\rm rad}$ monotonically decreases from the stellar center, which defines a clear boundary of the convective core according to the Schwarzschild criterion. Our results of numerical calculations show that the overshooting mixing presents a behaviour of exponential decay similar with Herwig (2000), and the overshooting distance is to make $\nabla_{\rm rad} \simeq \nabla_{\rm ad}$ at the boundary of the convective core, in agreement with the prediction of the self-driving mechanism (Castellani, Giannone \& Renzini (1971a). 

In the middle stage, the radiative temperature gradient $\nabla_{\rm rad}$ shows a minimum value in the convective core, dividing the convection zone into two if the minimum value of $\nabla_{\rm rad}$ is smaller than the adiabatic temperature gradient $\nabla_{\rm ad}$. For the single-zone case, the overshooting mixing still shows an exponential decay behaviour, but the overshooting distance is much smaller than in the initial stage. For the double-zone case, the overshooting mixing is similar to that of the single-case beyond the convective core, while it almost stops on both sides of the above convective shell. Therefore, the convective core continues to expand, until its boundary touches the inner edge of the above convective shell, results in the merge of the two convection zones into one convective core. This repeatedly suspensions of the overshooting mixing beyond the convective core is similar to the the maximal overshoot scheme of Constantino et al. (2015), which effectively restrains the convective core to expand too much large. 

In the final stage, continuous injection of fresh helium into the convective core by the overshooting mixing happens, which is similar to the "core breathing pulses" (Sweigart \& Demarque 1973; Castellani et al. 1985). There are two times of the helium injection taking place during this period, while typically one might expect about three times (e.g., Caloi 1989). 

In summary, the development of the convective core shows a complicated and variable behaviours for sdB stars. Various mechanisms proposed previously play a major role at different stages during the sdB evolution.

\acknowledgments 
This work was funded by the NSFC of China (Nos. 11333006, 11733008, and 11521303), and by Yunnan Province (No. 2017HC018). The authors thank Jie Su, Qian-sheng Zhang, Tao Wu, C. Aerts, and J. Christensen-Dalsgaard for their unreserved helps and fruitful discussions. Special thanks are given to an anonymous referee, whose comments helped us to significantly improve the manuscript. 
\software{MESA(v6596; Paxton et al. 2011, 2013, 2015, 2018)}

{}

\clearpage

\end{document}